\documentclass[reprint,nofootinbib,amsmath,amssymb,aps,floatfix]{revtex4-2}
\usepackage{graphicx}
\usepackage{dcolumn}
\usepackage[colorlinks,linkcolor=magenta,anchorcolor=cyan,citecolor=blue]{hyperref}
\usepackage{bm}
\usepackage[multiple]{footmisc}
\usepackage{appendix}
\usepackage{color}

\begin{document}
\title{Scalar Perturbation Around Rotating Regular Black Hole: Superradiance Instability and Quasinormal Modes}
\author{Zhen Li}
\email{zhen.li@nbi.ku.dk}
\affiliation {DARK, Niels Bohr Institute, University of Copenhagen, Jagtvej 128, 2200 Copenhagen Ø, Denmark}
\date{\today}

\begin{abstract}
Black holes provide a natural laboratory to study particle physics and astrophysics. When black holes are surrounded by matter fields, there will be plenty of phenomena which can have observational consequences, from which we can learn about the matter fields as well as black hole spacetime. In this work, we investigate the massive scalar field in the vicinity of a newly proposed rotating regular black hole inspired by quantum gravity. We will especially investigate how this non-singular spactime will affect the superradiance instability and quasinormal modes of the scalar filed. We derive the superradiant conditions and the amplification factor by using the Matching-asymptotic Method, and the quasinormal modes are computed through Continued Fraction Method. In the Kerr limit, the results are in excellent agreements with previous research. We also demonstrate how the quasinormal modes will change as a function of black hole spin, regularity described by a parameter $k$ and scalar field mass respectively, with other parameters taking specific values.
\end{abstract}
\maketitle

\section{introduction}

Our current best understanding on gravitational interaction is described by general relativity (GR). The recent observation of gravitational waves \cite{gw1,gw2,gw3} and black hole shadows\cite{shadow1,shadow2} provide even more evidences on this fascinating theory. However, GR also faces several challenges, such as, the incompatibility between GR and quantum theory \cite{cha1}, the singularities \cite{cha2,cha22}, the late time acceleration of the universe and so on \cite{de,de1,de2}. Among these, the singularities in classical GR are most severe. Because it is widely belief that singularities do not exist in nature, rather they reveal the limitations of GR. Therefore, the idea of regular black holes may provide a solution or a trial to the singularity problem. The regular black holes are the solutions that have horizons and non-singular at the origin, and their curvature invariants are regular everywhere\cite{re,re1,re2,re3,re4}. A novel spherical symmetric regular black hole proposed in \cite{nre1,nre2,nre3} and reformulated in \cite{nre4} is a very promising solution to the singularity problem. Later it has also been generalized to the rotating axisymmetric scenario\cite{nre5,nre6,nre7}. The exponential convergence factor is used in these regular black holes, which is also used in formulation of the quantum gravity\cite{qg}.

Scalar filed play a crucial role in the fundamental physics as well as astrophysics, like the inflation field \cite{inf,inf1,inf2} and also in the dark energy models\cite{dde}. Dark matter could also be a kind of scalar field, especially, the ultralight scalar field dark matter could have some advantages over the standard Lambda cold dark matter model\cite{dm}. When the Compton wavelength of the scalar field particles are comparable to the characteristic size of the black hole horizon, they can efficiently extract rotational energy from rotating black holes through superradiance instabilities and form macroscopic quasinormal condensates\cite{sr,sr1}. This provide a unique way and natural laboratory to detect the ultralight scalar field particles through black hole observations, for example, they will leave imprints on the gravitational waves\cite{srgw, srgw1}. Because of this and its importance in black hole physics, superradiance recently attracts plenty of attention from science community, and physicists have performed investigation in many different aspects and scenarios\cite{sr2,sr3,sr4,sr5,sr6,sr7,sr8,sr9,sr10,sr11,sr12,sr13,sr14,sr15,sr16,sr17,sr18}. It is also worth to mention that there are alternative mechanisms for energy extraction from a rotating black hole, such as Penrose process\cite{asr1,asr11}, the Blandford-Znajek process\cite{asr2}, magnetic reconnection process\cite{asr3,asr4,asr5} and so on, which may also produce (charged) scalar field particles.

Thus, to study the phenomenology of scalar field around rotating regular black holes will provide us much more insights on both gravity, astrophysics and particle physics. Usually, the scalar filed will be taken as a test field or perturbation filed such that it will not shift the black hole background spacetime. There are some related works on this topic but with different focus or regular spacetime\cite{rlw,rlw1,rlw2}. In this work, we will study the superradiance instabilities and quasinormal modes of scalar field around the newly proposed rotating regular black hole\cite{nre5,nre6,nre7}. We will demonstrate how the regular parameter affects the superradiance and quasinormal modes. 

The structure of this paper is as follows: In Section.\ref{sec2}, we will introduce the rotating regular black hole spacetime. In Section.\ref{sec3}, we will solve the massive Klein-Gordon equation in this spacetime, and obtained the radial and angular equations. Then, in Section.\ref{sec4}, we will analysis the superradiance instabilities and compute the amplification factor. Then, in Section.\ref{sec5}, we will compute the quasinormal modes by Continued Fraction method, we also demonstrate how the quasinormal modes will change as a function of black hole spin, regular parameter and scalar field mass respectively. In Section.\ref{sec6}, we will make a conclusion and discussion.

\section{rotating regular black hole}\label{sec2}
The metric of non-singular rotating black hole mentioned in the introduction could be written in the  Boyer–Lindquist coordinates as \cite{nre5,nre6,nre7}, 
\begin{equation}\label{metric}
\begin{aligned}
\mathrm{d} s^{2}=&-\left(1-\frac{2 M r \mathrm{e}^{-k / r}}{\Sigma}\right) \mathrm{d} t^{2}+\frac{\Sigma}{\Delta} \mathrm{d} r^{2}+\Sigma \mathrm{d} \theta^{2} \\
&-\frac{4 a M r \mathrm{e}^{-k / r}}{\Sigma} \sin ^{2} \theta \mathrm{d} t \mathrm{~d} \phi \\
&+\left[r^{2}+a^{2}+\frac{2 M r a^{2} \mathrm{e}^{-k / r}}{\Sigma} \sin ^{2} \theta\right] \sin ^{2} \theta \mathrm{d} \phi^{2}
\end{aligned}
\end{equation}
with $\Sigma=r^{2}+a^{2} \cos ^{2} \theta, \Delta=r^{2}+a^{2}-2 M re^{-k / r}$. and $M$, $a$, and $k$ are three parameters, which were assumed to be positive. The Kerr metric could be reduced when set $k/r=0$. 

To show the regularity of this metric, it is convenient to study the spacetime invariants, for example, the Kretschmann invariant $\mathrm{K}=R_{a b c d} R^{a b c d}\left(R_{a b c d}\right.$ is the Riemann tensor).
\begin{equation}
K=\frac{4 M^{2} \mathrm{e}^{\frac{-2 k}{r}}}{r^{6} \Sigma^{6}}\left(\Sigma^{4} k^{4}-8 r^{3} \Sigma^{3} k^{3}+A k^{2}+B k+C\right)
\end{equation}
where $A$, $B$, and $C$ are functions of $r$ and $\theta$, given by
\begin{equation}
\begin{array}{l}
A=-24 r^{4} \Sigma\left(-r^{4}+a^{4} \cos ^{4} \theta\right) \\
B=-24 r^{5}\left(r^{6}+a^{6} \cos ^{6} \theta-5 r^{2} a^{2} \cos ^{2} \theta \Sigma\right) \\
C=12 r^{6}\left(r^{6}-a^{6} \cos ^{6} \theta\right)\\
\quad\quad-180 r^{8} a^{2} \cos ^{2} \theta\left(r^{2}-a^{2} \cos ^{2} \theta\right)
\end{array}
\end{equation}
For $M \neq 0$, they are regular everywhere.

The solutions of equation 
\begin{equation}\label{eh}
\Delta=r^{2}+a^{2}-2 M re^{-k / r}=0
\end{equation}
will give us the event horizons. The numerical results of horizon structure with different parameters were discussed in \cite{nre2}. However, there are no analytical solutions.

Despite this, we can use approximation method to solve (\ref{eh}) analytically as long as $k/M\ll1$, and it also satisfies the condition for (\ref{eh}) to have two distinct real solutions (see \cite{nre2}), i.e, less than the critical value $k_c^{EH}$ which decreases with the increase in $a$, for $a=0.9M$, $k_c^{EH}\approx 0.1M$, for $a=0.95M$, $k_c^{EH}\approx 0.05M$. In the Kerr limit, $\Delta_{kerr}=r^{2}+a^{2}-2Mr=(r-r_+)(r-r_-)$, where $r_+$ and $r_-$ are called event and inner horizon of Kerr black hole respectively. They can be seen as the zeroth order (with respect to $k/r$) solution to equation (\ref{eh}). Because the equation (\ref{eh}) can be written as
\begin{equation}\label{newapp}
r^{2}+a^{2}-2Mr=2 M r(e^{-k/r}-1)
\end{equation}
where the right-hand side is much smaller than the left-hand side if $k/r \ll 1$, so the right-hand side is the small perturbation. Therefore, if we brought the zeroth order solutions $r_{\pm}$ into the right-hand side of (\ref{newapp}), we will get high order approximation solutions, there are
\begin{equation}
\begin{array}{l}
\Delta_{kerr}-2 M r_{+}(e^{-k/r_{+}}-1)=(r-{r}_+^I)(r-\tilde{r}_-)\\
\Delta_{kerr}-2 M r_{-}(e^{-k/r_{-}}-1)=(r-\tilde{r}_+)(r-{r}_-^I)
\end{array} 
\end{equation}
where ${r}_+^I$ and ${r}_-^I$ could be seen as the first order approximate solutions to (\ref{eh}), i.e,
\begin{equation}\label{eh1}
\Delta\approx(r-{r}_+^I)(r-{r}_-^I)
\end{equation}
where $\tilde{r}_+$ and  $\tilde{r}_-$ are the two extra roots because we are solving two quadratic equations, and they are numerically less accurate compared to ${r}_+^I$ and ${r}_-^I$. The explicit forms for ${r}_+^I$ and ${r}_-^I$ are given by
\begin{align}
{r}_+^I=M +\sqrt{M^{2}-a^{2}+2 M r_{+}(e^{-k/r_{+}}-1)}\\
{r}_-^I=M -\sqrt{M^{2} -a^{2}+2 M r_{-}(e^{-k/r_{-}}-1)}
\end{align}
For better accuracy, we can carry ${r}_+^I$ and ${r}_-^I$ back to the right-hand side of (\ref{newapp}) and repeat the process above to get more accurate second order solutions of (\ref{eh}).
\begin{align}
{r}_+^{II}=M +\sqrt{M^{2}-a^{2}+2 M {r}_+^I(e^{-k/ {r}_+^I}-1)}\\
{r}_-^{II}=M -\sqrt{M^{2} -a^{2}+2 M {r}_-^I(e^{-k/ {r}_-^I}-1)}
\end{align}
even third order solutions
\begin{align}\label{r++}
{r}_+^{III}=M +\sqrt{M^{2}-a^{2}+2 M {r}_+^{II}(e^{-k/ {r}_+^{II}}-1)}\\
{r}_-^{III}=M -\sqrt{M^{2} -a^{2}+2 M {r}_-^{II}(e^{-k/ {r}_-^{II}}-1)}\label{r--}
\end{align}
they could be seen as the event horizon and inner horizon of metric (\ref{metric}). One could repeat the approximation steps to get more higher order solutions, but third order ${r}_+^{III}$ and ${r}_-^{III}$ are sufficient in this work, see Appendix.\ref{ap1}. Here after we will define $\hat r_+ \equiv {r}_+^{III}$ and $\hat r_- \equiv {r}_-^{III}$ for simplicity.

\section{decoupled master equations for massive scalar field}\label{sec3}
The dynamics of a massive scalar field $\Phi$ in the spacetime (\ref{metric}) is governed by the Klein-Gordon equation 
\begin{equation}
\left(\nabla^{a} \nabla_{a}-\mu^{2}\right) \Phi=  (\sqrt{-g})^{-1} \partial_{\mu}\left( \sqrt{-g} g^{\mu \nu} \partial_{\nu} \Phi\right)-\mu^{2} \Phi=0
\end{equation}
where $g=det(g_{\mu\nu})$ and $\mu$ is the mass of the scalar field. We can rewrite it more explicitly in the Boyer–Lindquist coordinates as
\begin{align}\label{master}
&\left(\frac{\left(r^{2}+a^{2}\right)^{2}}{\Delta}-a^{2} \sin ^{2} \theta\right) \partial_{t} \partial_{t} \Phi+\frac{4 M a re^{-k / r}}{\Delta} \partial_{t} \partial_{\phi} \Phi \nonumber\\
&+\left(\frac{a^{2}}{\Delta}-\frac{1}{\sin ^{2} \theta}\right) \partial_{\phi} \partial_{\phi} \Phi-\partial_{r}\left(\Delta \partial_{r} \Phi\right)  \nonumber\\
&-\frac{1}{\sin \theta} \partial_{\theta}\left(\sin \theta \partial_{\theta} \Phi\right)+\mu^{2} \Sigma \Phi=0
\end{align}
For the axisymmetric and asymptotically flat black-hole spacetimes, the test Klein-Gordon allows for the separation of variables\cite{sep}. Since the spacetime symmetry and asymptotic behavior of Kerr black hole also apply to rotating regular black hole (\ref{metric}) as well\cite{nre2}, so we can decompose the field with the ansatz 
\begin{equation}\label{ansa}
\Phi\left(x^{\mu}\right)=e^{-i \omega t} e^{i m \phi}S_{l m}(\theta) R_{l m}(r)
\end{equation}
where $\omega$ is the frequency and it is permitted to be complex. The sign of $Im(\omega)$ determines whether the solution is decaying $(Im(\omega)< 0)$ or growing $(Im(\omega)> 0)$ in time. Carrying (\ref{ansa}) to equation (\ref{master}), this leads to two ordinary differential equations, also called the Teukolsky equations \cite{tes}. For the radial part, 
\begin{equation}\label{radial}
\begin{aligned}
\frac{d}{d r}\left(\Delta \frac{d R_{l m}}{d r}\right) &+\left(\frac{\omega^{2}\left(r^{2}+a^{2}\right)^{2}-4 M a m\omega re^{-k / r}  +m^{2} a^{2}}{\Delta}\right.\\
&\left.-\left(\omega^{2} a^{2}+\mu^{2} r^{2}+\Lambda_{l m}\right)\right) R_{l m}(r)=0
\end{aligned}
\end{equation}
where $\Lambda_{l m}$ is the separation constant, they are the eigenvalues with respect to the following angular part equation,
\begin{align}\label{angular}
\frac{1}{\sin \theta} \frac{d}{d \theta}\left(\sin \theta \frac{d S_{l m}}{d \theta}\right)&+\left(a^{2}\left(\omega^{2}-\mu^{2}\right) \cos ^{2} \theta\right. \nonumber\\
&\left.-\frac{m^{2}}{\sin ^{2} \theta}+\Lambda_{l m}\right) S_{l m}(\theta)=0
\end{align}
The angular solutions $S_{l m}(\theta)$ are spheroidal harmonics $S_{l m}=S_{l}^{m}(\cos \theta ; c)$. In the non-rotating limit, the spheroidal harmonics reduce to spherical harmonics $Y_{l m}$ and $\Lambda_{l m}\approx l(l+1)$.

We define $u(r) \equiv \sqrt{r^{2}+a^{2}} R_{l m}(r)$ and switch to the tortoise coordinate via $d r_{*}=\frac{r^{2}+a^{2}}{\Delta} d r$, after some algebra, the radial function (\ref{radial}) takes the following Schrodinger-like form
\begin{equation}\label{u}
\frac{d^{2} u\left(r_*\right)}{d r_*^ 2}+\mathcal{V}(r) u\left(r*\right)=0
\end{equation}
with the effective potential $\mathcal{V}(r)$ given by
\begin{widetext}
\begin{align}\label{po}
\mathcal{V}(r)=&\left(\omega-\frac{a m}{a^{2}+r^{2}}\right)^{2}-\frac{\Delta }{({a^{2}+r^{2}})^2}\left ( a^2\omega^2+\mu ^2r^2+\Lambda _{lm}-2am\omega \right )\nonumber\\
&- \frac{r^2\Delta ^2}{({a^{2}+r^{2}})^4} -\frac{\Delta+2r^2-2Me^{-k/r}(k+r) }{({a^{2}+r^{2}})^2} +\frac{4r^2\Delta }{({a^{2}+r^{2}})^3}
\end{align}
\end{widetext}
the first line comes form the potential of equation (\ref{radial}) divided by $(r^2 + a^2)^2$. The second line represents the effect of introducing the tortoise coordinate $dr_{*}$.

\section{superradiance instability}\label{sec4}
The incident scalar waves could be amplified when scattered off of a rotating or charged black hole, within certain parameter space of the black hole. This is so called Superradiance.

\subsection{superradiance modes}
In this section, we will study the conditions for the happening of superradiance. Now we consider the following asymptotic behaviour of the solutions or boundary conditions of equation (\ref{u}),
\begin{equation}\label{sb}
\begin{aligned}
u_{h}(r_*) &=\mathcal{A}_{\mathcal{T}} \exp \left(-i k_{h} r_{*}\right), r_{*} \longrightarrow -\infty(r \rightarrow \hat{r}_{+}) \\
u_{\infty}(r_*) &=\mathcal{A}_{\mathcal{I}} \exp \left(-i k_{\infty} r_{*}\right)\\
&+\mathcal{A}_{\mathcal{R}}\exp \left(i k_{\infty} r_{*}\right),  r_{*} \longrightarrow \infty(r \rightarrow \infty)
\end{aligned}
\end{equation}
where $k_{h}=\sqrt{\mathcal{V}\left(r \rightarrow \hat{r}_{+}\right)}=\omega-m\Omega_{h}$,  $k_{\infty}=\sqrt{\mathcal{V}(r \rightarrow \infty)}=\sqrt{\omega^{2}-\mu^{2}}$.
These boundary conditions describe an incoming wave from spatial infinity with an amplitude of $\mathcal{A}_{\mathcal{I}}$, which scatters off the event horizon and produces reflected and transferred waves with amplitudes of $\mathcal{A}_{\mathcal{R}}$ and $\mathcal{A}_{\mathcal{T}}$, respectively.

Now, by equating the Wronskian quantity $$W=\left(u \frac{d u^{*}}{d r_{*}}-u^{*} \frac{d u}{d r_{*}}\right)$$ for regions near the event horizon with its counterparts at infinity, we can get
\begin{equation}
\left|\mathcal{A}_{\mathcal{I}}\right|^{2}-\left|\mathcal{A}_{\mathcal{R}}\right|^{2}=\frac{\omega-m \Omega_{h}}{\sqrt{\omega^{2}-\mu^{2}}}\left|\mathcal{A}_{\mathcal{T}}\right|^{2}
\end{equation}
where $\Omega_h=a/(\hat{r}_+^2+a^2)$, According to the above equation, for superradiance to occur, the amplitude of the reflected waves must be greater than the amplitude of the incident waves, and the following frequency criteria must be met
\begin{equation}\label{cds}
\mu<\omega<m\Omega_{h}
\end{equation}
The frequency or modes satisfying above condition is called superradiance modes.

\subsection{amplification factor}
The degree of amplification caused by the superradiance is described by the Amplification Factor, it can be computed via
\begin{equation}\label{aff}
Z_{l m}=\frac{\left|\mathcal{A}_{\mathcal{R}}\right|^{2}}{\left|\mathcal{A}_{\mathcal{I}}\right|^{2}}-1
\end{equation}
Since the Kerr spacetime's symmetry and asymptotic behavior also applys to rotating regular black hole (\ref{metric}) as well, the derivation of this section is essentially similar to that of Kerr \cite{sr}, with the exception that $r_+$ is replaced by $\hat{r}_ +$. However, the geometry of near-horizon region is altered as a result of this difference, which also results in a different amplification factor. Just to be self-content, we will briefly review the calculation steps of amplification factor $Z_{l m}$. 

We assume that the Compton wavelength of scalar field particle is significantly greater than the black hole gravitational size or $\mu M \ll 1$, we also consider the low-frequency regime $ \omega M \ll 1$ which also implies $ a \omega \ll 1$, these conditions allow us to use the matching-asymptotic techniques \cite{match,match1} as follow.

We divide the space into two overlapping regions, i.e, the near-region $\omega (r- r_+) \ll 1$, and the far-region $ r-r_+\gg M$. We will solve the radial equation (\ref{radial}) at these two regions and then match them in their overlapping region, this will give us the analytical solutions to the amplitudes, so that we can compute the amplification factor.

The radial equation (\ref{radial}) can be written as
\begin{widetext}
\begin{align}\label{xradial}
x^2(1+x)^{2} \frac{d^{2} R_{lm}}{d x^{2}}&+ x(x+1)(2 x+1) \frac{d R_{lm}}{d x}+(\beta^{2} x^{4}-(\omega^{2} a^{2}+\Lambda_{l m}) x(x+1)\nonumber\\
&-\mu^{2} ((\hat r_+-\hat r_-)x+\hat r_+)^2x(x+1)+Q^{2}) R_{lm}=0
\end{align}
\end{widetext}
where we defined new variables 
\begin{align}
x&=\frac{r-\hat r_+}{\hat r_+-\hat r_-} \\
\beta &=\omega(\hat r_+-\hat r_-)\\
Q&=\frac{\hat{r}_{+}^{2}+a^{2}}{\hat{r}_{+}-\hat{r}_{-}}(m \Omega_h-\omega)
\end{align}

In the near-region, we have $kx \ll 1 $ and $\mu^{2} ((\hat r_+-\hat r_-)x+\hat r_+)^2\approx \mu^{2}\hat r_+^2 $, such that equation (\ref{xradial}) is then approximately take forms as
\begin{align}
x^{2}(x+1)^{2} \frac{d^{2}R_{lm}}{d x^{2}}&+x(x+1)(2 x+1) \frac{d R_{lm}}{d x}\nonumber\\
&+\left(Q^{2}-l(l+1) x(x+1)\right) R_{lm}=0
\end{align}
the general solution satisfying the boundary condition (\ref{sb}) to the above equation is given by the hypergeometric functions
\begin{align}
R_{lm}=A_{1} (\frac{x+1}{x})^{i Q} F(-l, l+1, 1-2 i Q,-x)
\end{align}
the large $x$ behavior of above solution is
\begin{align}\label{x}
R_{lm} &\sim A_{1}x^{l} \frac{\Gamma(1-2 i Q) \Gamma(2l+1)}{\Gamma(1+l-2 i Q) \Gamma(l+1)}\nonumber\\
&+A_{1}x^{-l-1} \frac{\Gamma(1-2 i Q) \Gamma(-2l-1)}{\Gamma(-l) \Gamma(-l-2 i Q)}
\end{align}

In the far-region, equivalently $x\rightarrow \infty$, equation (\ref{xradial}) approximately give us
\begin{equation}
\frac{d^{2}{R_{lm}}}{d x^{2}}+\frac{2}{x} \frac{{d}{R_{lm}}}{{d} x}+\left(\xi^{2}-\frac{l(l+1)}{x^{2}}\right) {R_{lm}}=0
\end{equation}
where $\xi=(\hat{r}_{+}-\hat{r}_{-})\sqrt{\omega^2-\mu^2}$. The solution of this equation can be written in terms of the confluent hypergeometric function
\begin{align}\label{kk}
{R}_{lm}&=\exp (-i \xi x)C_{1} x^{l} U(l+1,2 l+2,2 i \xi x)\nonumber\\
&+\exp (-i \xi x)C_{2} x^{-l-1} U(-l,-2 l, 2 i \xi x)
\end{align}
Expanding it for small $kx \ll 1$, we obtain
\begin{equation}\label{k}
{R}_{lm}\sim C_{1} x^{l}+C_{2} x^{-l-1}
\end{equation}
Now, matching (\ref{x}) and (\ref{k}), we can get 
\begin{align}
C_{1}=A_{1} \frac{\Gamma(1-2 i Q) \Gamma(2 l+1)}{\Gamma(l+1) \Gamma(l+1-2 i Q)} \nonumber\\
C_{2}=A_{1} \frac{\Gamma(1-2 i Q) \Gamma(-1-2 l)}{\Gamma(-l-2 i Q) \Gamma(-l)} \nonumber
\end{align}

When $r\rightarrow \infty$, from (\ref{sb}), we can know the solution of (\ref{radial}) takes form as
\begin{equation}\label{rin}
{R}_{lm}\sim \frac{u_{\infty}(r_*)}{r} \sim \mathcal{A}_{\mathcal{I}} \frac{\exp \left(-i k_{\infty} r_{*}\right)}{r}+\mathcal{A}_{\mathcal{R}}\frac{\exp \left(i k_{\infty} r_{*}\right)}{r}
\end{equation}
Expanding (\ref{kk}) at infinity and matching to (\ref{rin}), we obtain the analytical expression for $\mathcal{A}_{\mathcal{I}}$ and $\mathcal{A}_{\mathcal{R}}$
\begin{align}
\mathcal{A}_{\mathcal{I}} &=C_{1} \frac{(-2 i)^{-l-1} \xi^{-l} \Gamma(2 l+2)}{k_{\infty} \Gamma(l+1)}+C_{2} \frac{(-2 i)^{l} \xi^{l+1} \Gamma(-2 l)}{k_{\infty}\Gamma(-l)}, \\
\mathcal{A}_{\mathcal{R}} &=C_{1} \frac{(2 i)^{-l-1} \xi^{-l} \Gamma(2 l+2)}{k_{\infty}\Gamma(l+1)}+C_{2} \frac{(2 i)^{l} \xi^{l+1} \Gamma(-2 l)}{k_{\infty}\Gamma(-l)}.
\end{align}

\begin{figure}[htbp]
\centering
\includegraphics[scale=0.6]{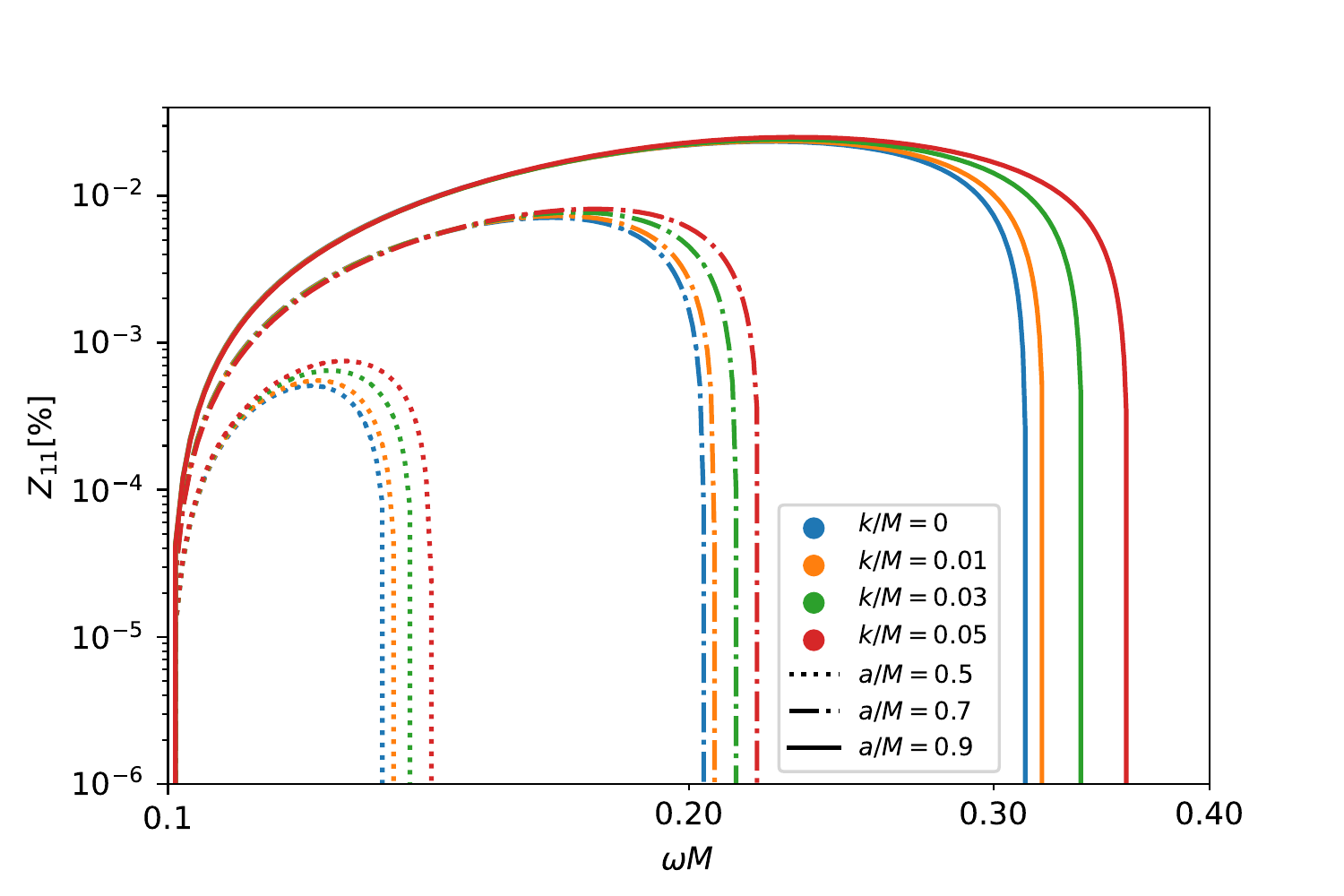}
\caption{The amplification factor $Z_{11}$ for $l=m=1$, $\mu M=0.1$, with three black hole spin $a=0.5M$, $a=0.7M$ and $a=0.9M$, and different regular parameter values $k =0.01M$, $k =0.03M$, $k =0.05M$, with the Kerr case ($k =0$) as reference.}
\label{af}
\end{figure}

After some algebra, we finally find the amplification factor (\ref{aff}) takes explicit form as
\begin{equation}\label{ll}
Z_{lm}=4 Q \xi^{2 l+1}\frac{(l !)^{4}}{((2 l) !)^{2}((2 l+1) ! !)^{2}} \\
\times \prod_{n=1}^l\left(1+\frac{4 Q^{2}}{n^{2}}\right)
\end{equation}
The formulas above are valid for any spin $a \le M $ provided $\mu M< \omega M \ll 1$. In Fig.\ref{af}, we plot $Z_{11}$ for different values of regular parameter $k$ and black hole spin, by setting $\mu M=0.1$. We can clearly see that the amplification starts when $\omega M > \mu M $, and dies out when it close to the threshold frequencies $m\Omega_h$. The amplification increases along with the black hole spin, and the parameter $k/M$ only affects the amplification when the frequencies close to the threshold $m\Omega_h$, and bigger $k/M$ will cause a bigger threshold frequency.

\section{quasinormal Modes}\label{sec5}
Quasinormal modes are solutions of the wave equation (\ref{u}), satisfying the following boundary conditions,
\begin{align}\label{b2}
u_{h}(r_*) &= \exp \left(-i k_{h} r_{*}\right),\quad r_{*} \longrightarrow -\infty(r \rightarrow \hat{r}_+) \nonumber\\
u_{\infty}(r_*) &=\exp \left(i k_{\infty} r_{*}\right), \quad r_{*} \longrightarrow \infty(r \rightarrow \infty)
\end{align}
which means there are only ingoing waves at the event horizon, while pure outgoing wave at spatial infinity. This condition leads to a discrete eigenvalue of frequencies. Quasinormal modes were referred to as the "fingerprints" of black holes. Because they are determined by the parameters of black holes, like mass and spin etc.

There are many methods to compute the quasinormal modes, see the reviews\cite{qnm,qnm1,qnm2}. In this section, we will use the popular Continued Fraction Method to compute the quasinormal modes, and this method has been used in many outstanding works even recently\cite{cfm,cfm1,cfm2}.

\subsection{Continued Fraction Method}
According to the boundary conditions (\ref{b2}), we can obtain a series solution to the radial equation (\ref{radial}), by setting $R_{l m}(r)$ as 
\begin{equation}\label{se1}
R_{\ell m}(r)=\left(r-\hat r_{+}\right)^{-i \sigma_{+}}\left(r-\hat r_{-}\right)^{i \sigma_{-}}y(r-\hat r_-)
\end{equation}
where $-i\sigma_{+}$ and $i\sigma_{-}$ are the indices of $R_{\ell m}(r)$ at singular points $r=\hat r_+$ and $r=\hat r_-$, they are given by
\begin{align}
\sigma_{+}=\frac{\omega \hat r_+ - am}{b}\\
\sigma_{-}=\frac{\omega \hat r_- - am}{b}
\end{align}
where $b=\hat r_+-\hat r_-$. For simplicity, we first define $x=r-\hat r_-$, now we have $\Delta=x(x-b)$. Next, we rewrite (\ref{se1}) and (\ref{radial}) in terms of $x$, and also substitute the series solution (\ref{se1}) into (\ref{radial}). Then we calculate the first term (derivatives) of (\ref{radial}). When the above operations were done, we collect all the terms with the same order of $y(x)$, $\frac{d y}{d x}$, $\frac{d^{2} y}{d x^{2}}$ respectively. At the end, we can get the second order differential equation for $y(x)$,
\begin{align}
x\left(x-b\right) \frac{d^{2} y}{d x^{2}}&+\left(B_{1}+B_{2} x\right) \frac{d y}{d x}+\left((\omega^{2}-\mu^2) x\left(x-b\right) \right. \nonumber\\
&\left.-2 \eta \sqrt{\omega^2-\mu^2}\left(x-b\right)+B_{3}\right) y=0
\end{align}
where $B_1$, $B_2$, $B_3$ and $\eta$ are given by
\begin{align}
 B_1 &= (-1-2i\sigma_{-})b\nonumber\\
 B_2 &= 2(i\sigma_{-}-i\sigma_{+}+1)\nonumber\\
 B_3 &= 2\omega^2\hat r_+^2+\omega^2(\hat r_++\hat r_-)^2+a^2\omega^2-\Lambda_{l m}-\mu^2\hat r_+^2\nonumber\\
 &\quad+i(\sigma_{-}-\sigma_{+})-(\sigma_{-}-\sigma_{+})^2\nonumber\\
 \eta &= -(\omega^2-\mu^2/2)(\hat r_+ + \hat r_-)/\sqrt{\omega^2-\mu^2}\nonumber
\end{align}
The function $y(x)$ can be expanded further in a power series
\begin{equation}\label{se2}
y(x)=e^{i k_{\infty} x} x^{-(1 / 2) B_{2}-i \eta} \sum_{n=0}^{\infty} d_{n}\left(\frac{x-b}{x}\right)^{n} 
\end{equation}
By substituting the above series solution $y(x)$ to $R_{\ell m}(r)$, and then to the radial equation (\ref{radial}). We can obtain the coefficients $d_n$ which satisfies a three term recurrence relation as follow
\begin{align}\label{cf1}
\alpha_{0} d_{1} +\beta_{0} d_{0}& =0 \\
\alpha_{n} d_{n+1}+\beta_{n}d_{n}+\gamma_{n}d_{n-1}& =0, \quad n=1,2,3,... \ldots\label{cf2}
\end{align}
where the coefficients are
\begin{align}
\begin{array}{l}
\alpha_{n}=n^{2}+\left(c_{0}+1\right) n+c_{0} \\
\beta_{n}=-2 n^{2}+\left(c_{1}+2\right) n+c_{3} \\
\gamma_{n}=n^{2}+\left(c_{2}-3\right) n+c_{4}-c_{2}+2
\end{array}
\end{align}
and the intermediate constant $c_n$ are defined as
\begin{equation}
\begin{array}{l}
c_{0}=B_{2}+B_{1} / b\\ c_{1}=-2\left(c_{0}+1+i\left(\eta-k_{\infty} b\right)\right) \\
c_{2}=c_{0}+2(1+i \eta) \\
c_{3}=-c_{4}-\frac{1}{2} B_{2}\left(\frac{1}{2} B_{2}-1\right)+\eta(i-\eta)+i k_{\infty} b c_{0}+B_{3} \\
c_{4}=\left(\frac{1}{2} B_{2}+i \eta\right)\left(\frac{1}{2} B_{2}+i \eta+1+B_{1} / b\right) \\
\end{array}
\end{equation}
If the series in (\ref{se1}) and (\ref{se2}) converges and the $r=\infty$ boundary condition (\ref{b2}) is satisfied, for a given $a$, $M$, $k$, $\mu$ and $\Lambda_{l m}$, the frequency $\omega$  must be a root of the continued fraction equation
\begin{equation}\label{cf}
0=\beta_{0}-\frac{\alpha_{0} \gamma_{1}}{\beta_{1}-} \frac{\alpha_{1} \gamma_{2}}{\beta_{2}-} \frac{\alpha_{2} \gamma_{3}}{\beta_{3}-} \ldots
\end{equation}
or any of its inversions. (\ref{cf}) is obtained by combining equation (\ref{cf1}) and (\ref{cf2}). The roots of (\ref{cf}) will give us the so called quasinormal modes. 

\subsection{Numerical Results}
For simplicity and consisting with the literature, we will choose units by setting $M = 1$ in the rest of this paper. Then the radial distance $r$, the regular parameter $k$ and black hole spin $a$ are measured in unit of $M$, while the frequency $\omega$ and scalar field mass $\mu$ are in unit of $M^{-1}$.

Our numerical procedures operate as follow, we first calculate the angular eigenvalues $\Lambda_{lm}$ using the Leaver method\cite{lea}, by fixing the values for ($k$, $l$, $m$, $a$, $\mu$). Then the continued fraction equation (\ref{cf}) depends only on the quasinormal frequency $\omega$.  For practical purposes, it is necessary to truncate the above continuing fraction to an order of $n$, We use a technique developed by Nollert \cite{nl} to approximate the value $n$. At the end, the root-finding algorithm (Built-in functions in \textit{Wolfram Mathematica}) will be applied to find the roots of the continued fraction equation (\ref{cf}). Previous calculations of quasinormal modes in the Kerr background \cite{valid1,valid2} are used to validate our numerical methods. The errors of quasinormal modes caused by using the approximation (\ref{r++}) and (\ref{r--}) are less than $10^{-2}$, see Appendix.\ref{ap1}, they become extreme accurate when the spin grows.

In Table.\ref{tab1} and Table.\ref{tab2}, we show some of the quasinormal frequencies for the fundamental mode with $(l=1,m=0)$ and $(l=1,m=1)$, by setting different black hole spins and regular parameter $k$. The scalar field mass has been set as $\mu=0.1$. In Table.\ref{tab3}, we show the dependency of quasinormal frequencies on the scalar field mass $\mu$ and regular parameter $k$, by setting $(l=1,m=1),a=0.5$. Please note that we used minus $Im(\omega)$ to represent the imaginary part of quasinormal modes here and after in this paper. In all three tables, the $k=0$ columns correspond to the quasinormal modes of Kerr black hole, which are in excellent six decimals agreements with the results obtained before by \cite{valid1,valid2}. 

\begin{figure}[htbp]
   \includegraphics[scale=0.6]{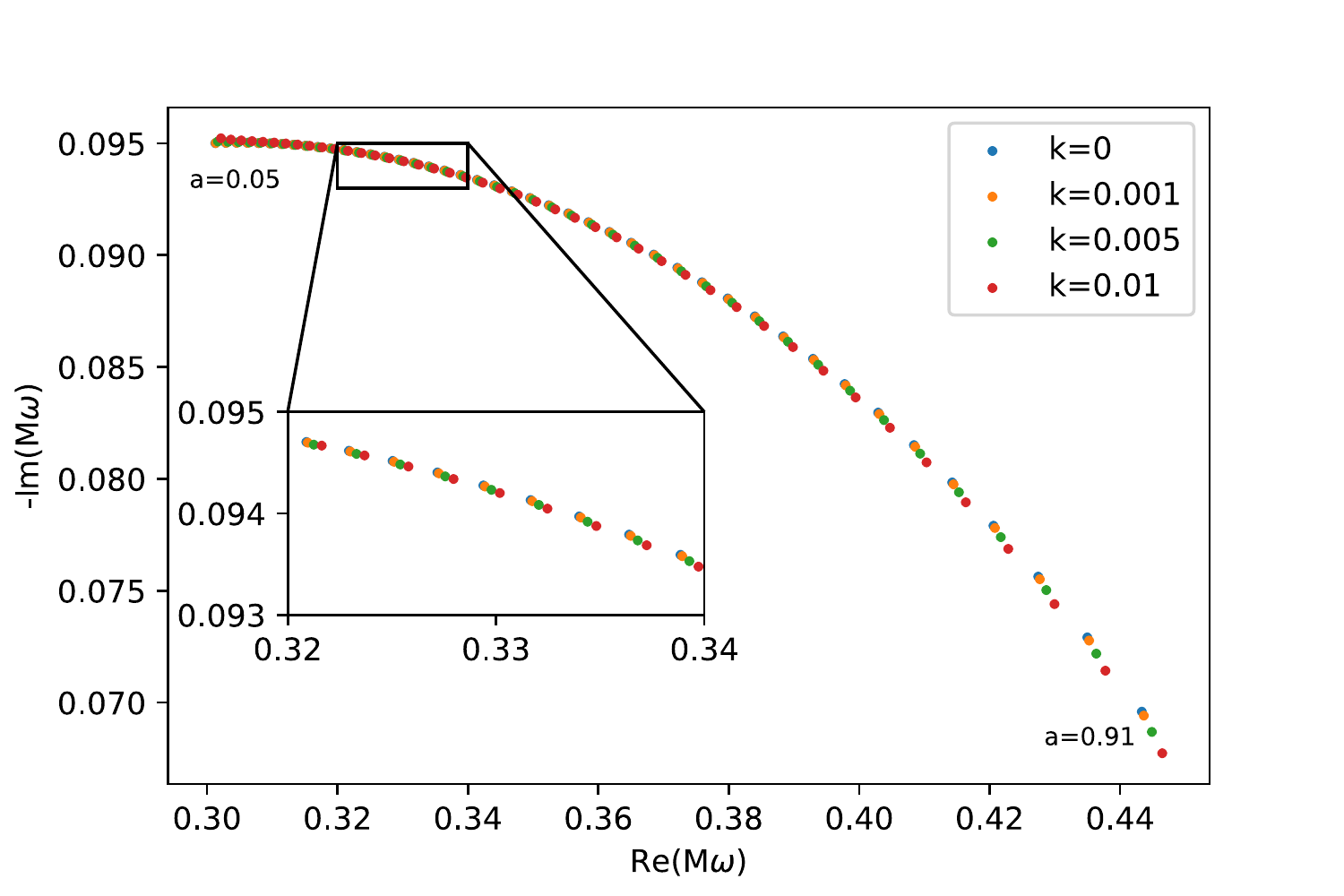}
  \caption{the $l=m=1$ fundamental quasinormal frequencies as a function of black hole spin (from $a= 0.05$ to $a= 0.91$ with spin intervals $\Delta a=0.02$), with the regular parameter $k=0, 0.001, 0.005, 0.01$, scalar filed mass $\mu=0.1$. }
\label{l1m1}
\end{figure}

From Table.\ref{tab1} and Table.\ref{tab2}, we notice that the real part of quasinormal frequencies grow along with the black spin, while the imaginary part decrease with spin. The regular parameter $k$ does play a role on the quasinormal frequencies. We also computed ($l=1$, $m=1$) quasinormal frequencies with smaller spin intervals $\Delta a=0.02$ from $a=0.05$ to $a=0.91$, with regular parameter $k=0, 0.001, 0.005, 0.01$, The results were plotted in Fig.\ref{l1m1}. We can see that the higher spin and bigger regular parameter $k$, the more deviation from Kerr black hole ($k=0$).

\begin{table*}[]
\setlength{\tabcolsep}{3mm}
\begin{tabular}{cllllllll}
\hline \hline
$\mu=0.1$ & \multicolumn{2}{c}{$k$ = 0}  & \multicolumn{2}{c}{$k$=0.001} & \multicolumn{2}{c}{$k$=0.005} & \multicolumn{2}{c}{$k$=0.01}                             \\
$a$    & Re($\omega$) & -Im($\omega$) & Re($\omega$) & -Im($\omega$) & Re($\omega$) & -Im($\omega$) & Re($\omega$) & -Im($\omega$) \\ \hline
0.1 & 0.297602 & 0.094884 & 0.297658 & 0.094884 & 0.297921 & 0.094914 & 0.298290 & 0.094981 \\
0.2 & 0.298164 & 0.094661 & 0.298218 & 0.094658 & 0.298452 & 0.094661 & 0.298771 & 0.094684 \\
0.3 & 0.299116 & 0.094273 & 0.299170 & 0.094269 & 0.299397 & 0.094259 & 0.299698 & 0.094260 \\
0.4 & 0.300478 & 0.093692 & 0.300534 & 0.093686 & 0.300762 & 0.093668 & 0.301058 & 0.093653 \\
0.5 & 0.302285 & 0.092873 & 0.302342 & 0.092865 & 0.302574 & 0.092838 & 0.302874 & 0.092809 \\
0.6 & 0.304579 & 0.091742 & 0.304638 & 0.091732 & 0.304878 & 0.091692 & 0.305185 & 0.091646 \\
0.7 & 0.307413 & 0.090182 & 0.307475 & 0.090168 & 0.307725 & 0.090110 & 0.308043 & 0.090041 \\
0.8 & 0.310836 & 0.087989 & 0.310900 & 0.087968 & 0.311160 & 0.087884 & 0.311489 & 0.087780 \\
0.9 & 0.314815 & 0.084810 & 0.314880 & 0.084778 & 0.315143 & 0.084654 & 0.315474 & 0.084498                  \\ \hline\hline
\end{tabular}
\caption{Values of the quasinormal frequencies for the fundamental mode, with $l =1$, $m = 0$, $\mu=0.1$ for different values of $k$, and spin $a$.}
\label{tab1}
\end{table*}

\begin{table*}[]
\setlength{\tabcolsep}{3mm}
\begin{tabular}{cllllllll}
\hline \hline
$\mu=0.1$ & \multicolumn{2}{c}{$k$ = 0}  & \multicolumn{2}{c}{$k$=0.001} & \multicolumn{2}{c}{$k$=0.005} & \multicolumn{2}{c}{$k$=0.01}                             \\
$a$    & Re($\omega$) & -Im($\omega$) & Re($\omega$) & -Im($\omega$) & Re($\omega$) & -Im($\omega$) & Re($\omega$) & -Im($\omega$) \\ \hline
0.1 & 0.305329 & 0.095029 & 0.305390 & 0.095027 & 0.305674 & 0.095054 & 0.30607  & 0.095118 \\
0.2 & 0.314119 & 0.094920 & 0.314184 & 0.094915 & 0.314460 & 0.094908 & 0.314833 & 0.094921 \\
0.3 & 0.323981 & 0.094569 & 0.324052 & 0.094561 & 0.324347 & 0.094536 & 0.324735 & 0.094518 \\
0.4 & 0.335181 & 0.093883 & 0.335261 & 0.093871 & 0.335590 & 0.093828 & 0.336015 & 0.093783 \\
0.5 & 0.348105 & 0.092714 & 0.348198 & 0.092696 & 0.348576 & 0.092630 & 0.349059 & 0.092552 \\
0.6 & 0.363345 & 0.090805 & 0.363456 & 0.090780 & 0.363904 & 0.090678 & 0.364474 & 0.090554 \\
0.7 & 0.381888 & 0.087678 & 0.382025 & 0.087637 & 0.382580 & 0.087474 & 0.383285 & 0.087271 \\
0.8 & 0.405606 & 0.082262 & 0.405790 & 0.082191 & 0.406531 & 0.081904 & 0.407473 & 0.081540 \\
0.9 & 0.439045 & 0.071342 & 0.439332 & 0.071183 & 0.440495 & 0.070533 & 0.441982 & 0.069685\\
\hline\hline
\end{tabular}
\caption{Values of the quasinormal frequencies for the fundamental mode, with $l =m = 1$, $\mu=0.1$ for different values of $k$, and spin $a$.}
\label{tab2}
\end{table*}

\begin{table*}[]
\setlength{\tabcolsep}{3mm}
\begin{tabular}{cllllllll}
\hline \hline
$a$=0.5 & \multicolumn{2}{c}{$k$ = 0}  & \multicolumn{2}{c}{$k$=0.001} & \multicolumn{2}{c}{$k$=0.005} & \multicolumn{2}{c}{$k$=0.01}                             \\
$\mu$    & Re($\omega$) & -Im($\omega$) & Re($\omega$) & -Im($\omega$) & Re($\omega$) & -Im($\omega$) & Re($\omega$) & -Im($\omega$) \\ \hline
0   & 0.344753 & 0.094395 & 0.344848 & 0.094375 & 0.345234 & 0.094301 & 0.345726 & 0.094214 \\
0.1 & 0.348105 & 0.092714 & 0.348198 & 0.092696 & 0.348576 & 0.092630 & 0.349059 & 0.092552 \\
0.2 & 0.358230 & 0.087478 & 0.358317 & 0.087466 & 0.358671 & 0.087423 & 0.359125 & 0.087375 \\
0.3 & 0.375284 & 0.078022 & 0.375362 & 0.078020 & 0.375679 & 0.078016 & 0.376086 & 0.078016 \\
0.4 & 0.399201 & 0.062970 & 0.399267 & 0.062982 & 0.399536 & 0.063031 & 0.399884 & 0.063099 \\
0.5 & 0.429036 & 0.040234 & 0.429096 & 0.040270 & 0.429336 & 0.040421 & 0.429639 & 0.040617\\
\hline\hline
\end{tabular}
\caption{Values of the quasinormal frequencies for the fundamental mode, with $l =m = 1$, $a=0.5$ for different values of $k$, and mass $\mu$.}
\label{tab3}
\end{table*}

\begin{figure}[htbp]
  \centering
   \includegraphics[scale=0.5]{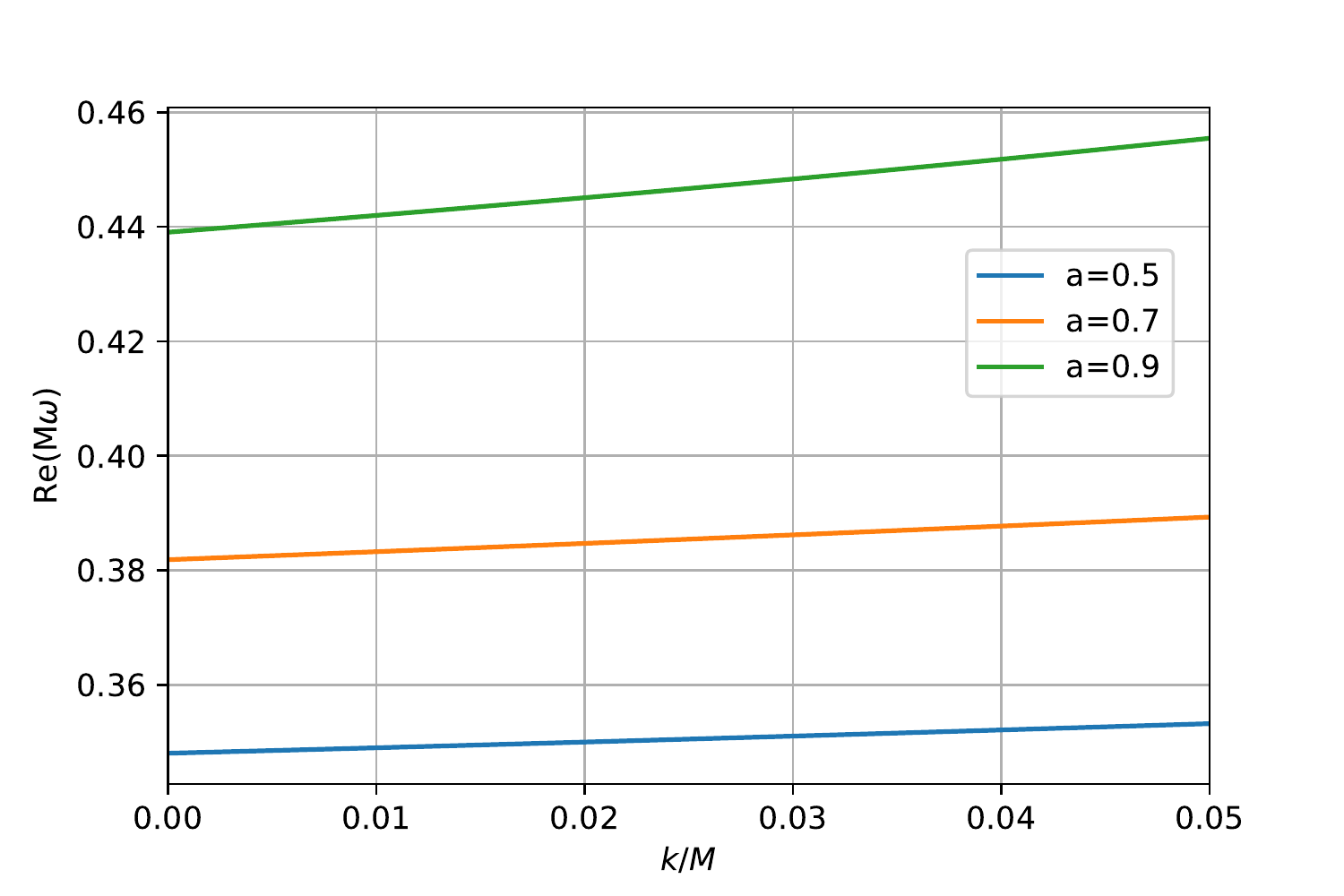}
    \includegraphics[scale=0.5]{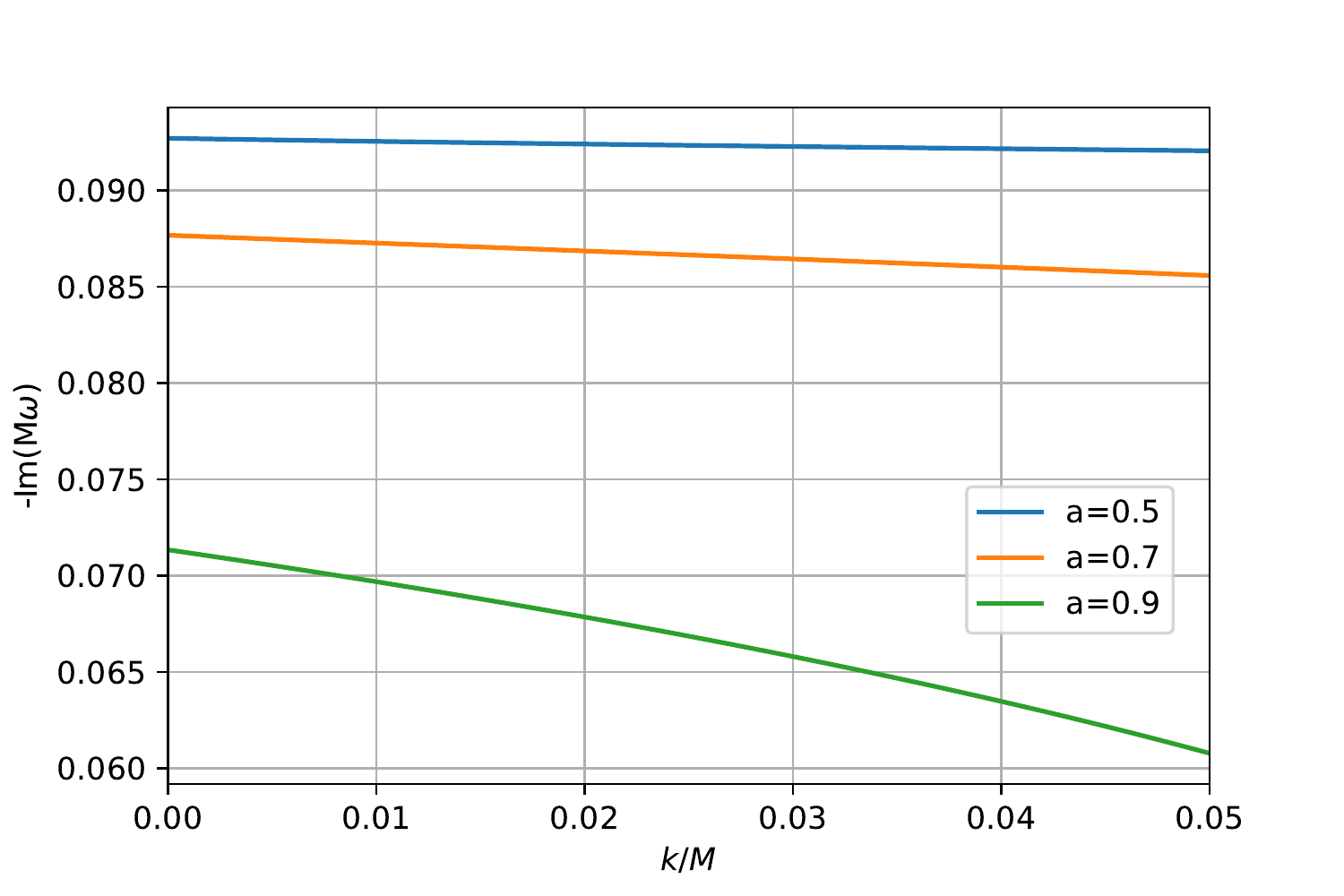}
  \caption{upper and lower plots are respectively the real and imaginary part of the fundamental quasinormal frequencies as a function of parameter $k$, by setting $l=m=1$, $\mu=0.1$ and three different spins $a=0.5, 0.7, 0.9$.}
\label{kx}
\end{figure}

To better show the dependency of quasinormal frequencies on the regular parameter $k$, we plot the real and imaginary part of quasinormal frequencies as a function of regular parameter $k$ in Fig.\ref{kx}, with three different high spins $(a\ge 0.5)$. We can see that regular parameter $k$ will increase the real part of quasinormal frequencies gently. For the imaginary part, the regular parameter $k$ decrease imaginary part of quasinormal frequencies, even more for higher spins. These features could provide us some insights on the connection bewteen the regularity of black hole and the stability of massive scalar field perturbation. All the imaginary part of quasinormal modes are negative (see Table.\ref{tab1}, \ref{tab2}, \ref{tab3} and Figure.\ref{l1m1}, \ref{kx}), which means the black hole  (\ref{metric}) is stable under massive scalar field perturbation. What's more, especially in the high spin regime, the increasing in regular parameter $k$ will cause a smaller imaginary part of quasinormal modes compare to Kerr black hole, which corresponds to a longer damping time. The regularity of black hole seems put more ‘elasticity’ onto the massive scalar field perturbation such that they will live longer than the Kerr black hole scenario. On the other hand, increasing in regular parameter $k$ will cause a bigger real part of quasinormal modes, which means it will increase the oscillation frequency of scalar field perturbation.
In addition, we again see black hole spins change quasinormal frequencies significantly for both real and imaginary part.

Last but not least, we investigate the quasinormal frequencies dependency on the scalar field mass $\mu$. From Table.\ref{tab3}, we can see that, for the real part, the value increase monotonously with the scalar field mass $\mu$. For the imaginary part, the value decrease monotonously with the scalar field mass $\mu$. It seems that, for $\mu \le 0.3$, the bigger regular parameter $k$, the bigger real part and smaller imaginary part. But for $\mu \ge 0.4$, it is not the case, where the bigger $k$, the bigger imaginary part. So, the scalar field mass will affect the relation between the regularity of black hole and the stability of massive scalar field perturbation. It maybe easy to understand because the different scalar field mass will significantly change the behavior of scalar field perturbation such that it will react differently to the regularity of black hole spacetime.

\section{conclusion and discussion}\label{sec6}
We have studied the massive scalar field perturbation around regular rotating black hole. We first introduced the newly proposed regular black hole spacetime metric (\ref{metric}) and some physical quantities, also we used approximation method to analytically solve the horizons. Then we separated and solved the massive Klein-Gordon equation in this spacetime and obtained the master equations(\ref{master}), also radial part (\ref{radial}) and angular part equations (\ref{angular}). With these equations, we studied the superradiance instability and quasinormal modes.

For the superradiance instability, we first discussed the conditions for the superradiance happen. The results show that the amplification happens when the frequencies are within certain parameter region (\ref{cds}). Then, we used the matching-asymptotic method to compute the amplification factor under small mass and low frequency approximations. At the end, we obtained a very net analytical expression for the amplification factor (\ref{ll}) and we plotted the $l=m=1$ modes with several parameters.

Regarding to the quasinormal modes, we applied the Continued Fraction Method to numerically calculated the quasinormal modes of the rotating regular black hole. We present three tables (Table.\ref{tab1}, \ref{tab2}, \ref{tab3}), and the $k=0$ columns could validate our numerical approaches. It is in excellent agreement with the previous results in the Kerr limit\cite{valid1,valid2}. The numerical results, in these tables also in Fig.\ref{kx}, are computed by selecting some parameters as the variables while others are settled down with certain values for the purpose of studying: (a) fundamental quasinormal modes as a function of black hole spin $a$; (b) fundamental quasinormal modes as a function of regular parameter $k$; (c) fundamental quasinormal modes as a function of scalar field mass $\mu$.

This work is the first step to study the perturbations around the non-singular rotating black hole (\ref{metric}). There are many researches could be conduct at the future. For example, the superradiance instability and quasinormal modes of the vector and gravitational perturbations (i.e, gravitational waves) in this spacetime. We believe it will give us more deeper understanding on black holes and gravity. We leave these research in the future.

\section*{acknowledgements}
The author would like to thank professor Steen Hansen for helpful discussions. The author also thanks the DARK cosmology centre at Niels Bohr Institute for supporting this research. This work is also financially supported by the China Scholarship Council.
\\
\\
\begin{appendix}
\section{the errors of approximation method}\label{ap1}

\begin{figure}[htbp]
  \centering
  \includegraphics[scale=0.5]{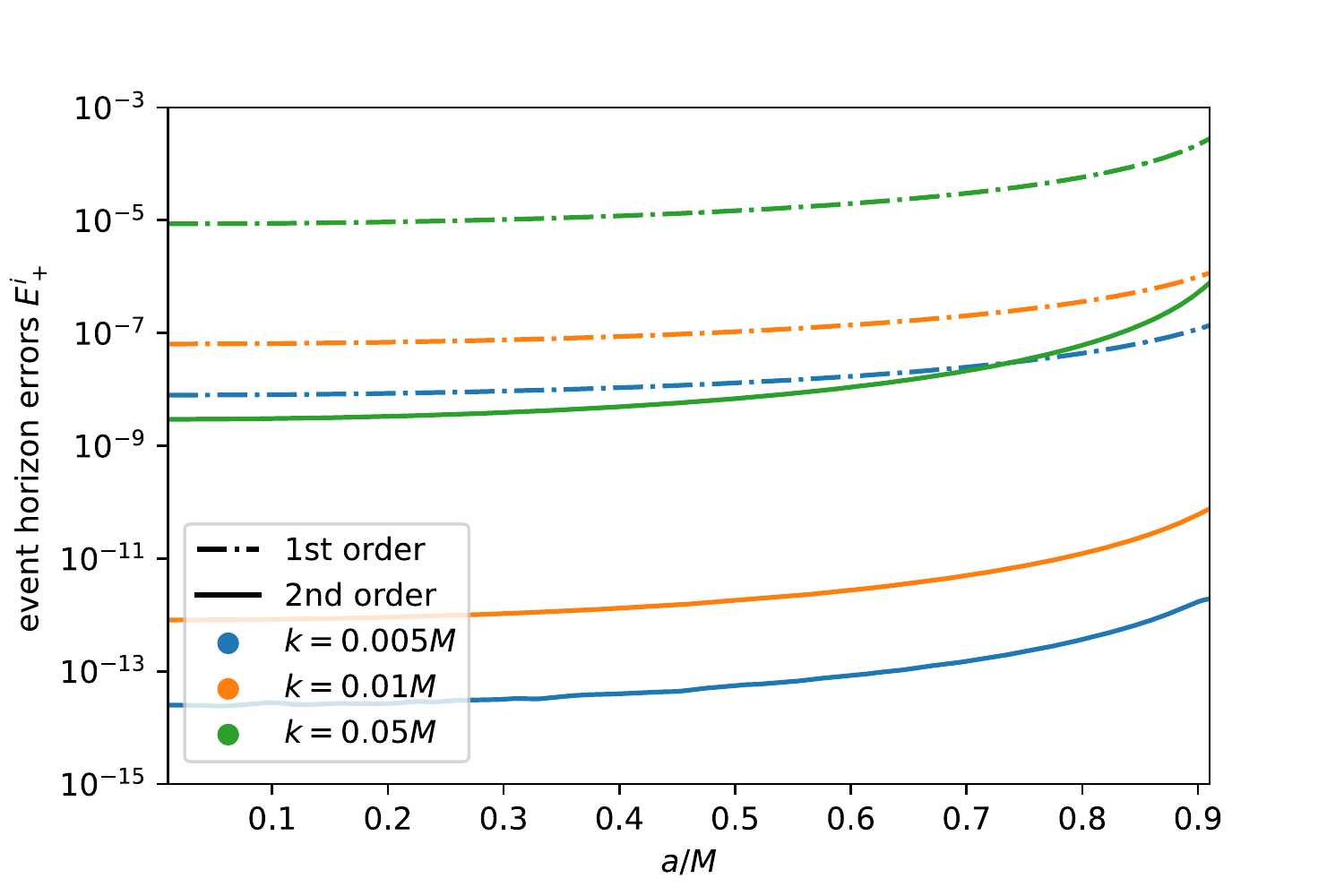}
   \includegraphics[scale=0.5]{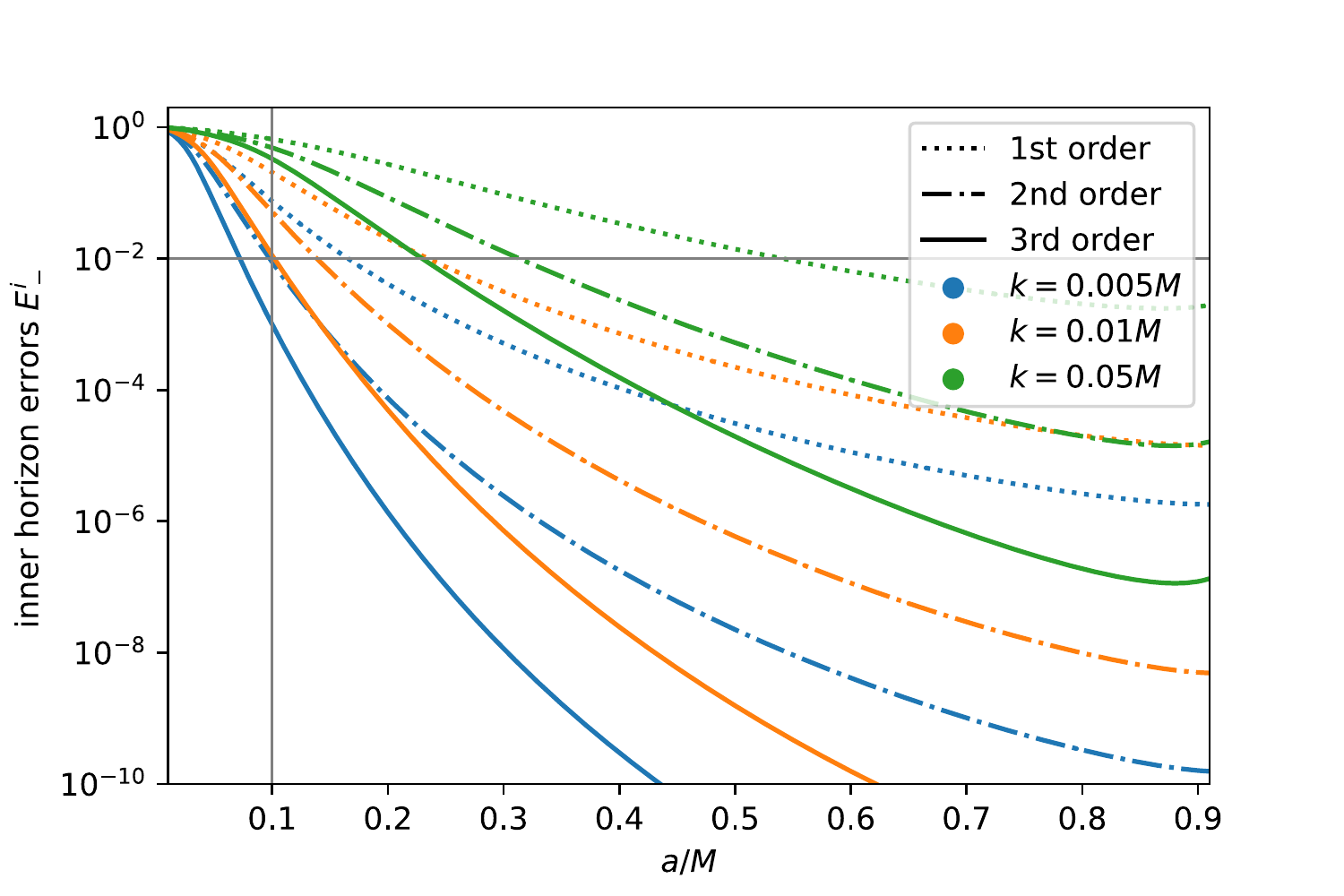}
  \caption{upper and lower plots are respectively the errors of high order event horizon and inner horizon, with regular parameter $k=0.005,0.01,0.05$ and black hole spin $a=0.01$ to $0.91$.}
\label{err}
\end{figure}

In Section.\ref{sec2}, we used the analytically approximation to solve the Delta function (\ref{eh}). To show the accuracy of this method, we plot the errors of high order solutions in Fig.\ref{err} as a function of black hole spin $a$, given different regular parameter $k$. We plot the errors up to the second order for event horizon and third order for inner horizon. The errors denoted as $E_{\pm}^i$ are computed by comparing to the numerical solutions $r_{\pm}^{num}$ of (\ref{eh}), i.e, 
\begin{equation}
E_{\pm}^i=\frac{\left| r_{\pm}^i - r_{\pm}^{num}\right|}{r_{\pm}^{num} }
\end{equation}
where $i=I,II,III$ represents the $i$-th order solution.

We can see from Fig.\ref{err}, the approximation errors drop down as the order goes up for both event and inner horizon, besides that, we can see the errors of event horizon are tiny $(\ll 10^{-3})$ even in the first order although they are slowly going up along with the black hole spin. However, the errors of inner horizon are significantly large at first order, but they drop down quickly as the order and black hole spin goes up.

From the order of magnitude, the errors of inner horizon are significantly larger than that of event horizons $E_{-}^i\gg E_{+}^i$ even in the high spin regime. Therefore, we shall focus on the inner horizon errors $E_{-}^{i}$, because it will dominate the errors of our approximation results. We plot a horizontal line $E_{-}^i=10^{-2}$ and a vertical line $a=0.1$. We would like to control the errors such that they always below $10^{-2}$ from $a=0.1$ to $0.9$. Then the regular parameter $k$ should be chosen not greater than $0.01M$, and third order solutions are sufficient for our goal.

The final equation to compute the amplification factor and quasinormal modes are (\ref{ll}) and (\ref{cf}) respectively, the essential quantities $Q$, $\xi$, $\alpha_n, \beta_n, \gamma_n$ for these equations are all proportional to $\hat r_+$, $\hat r_+$, $\omega$ or their quadratic. The accuracy of amplification factor and quasinormal modes, because of using the approximation solutions (\ref{r++}) and (\ref{r--}), are therefore in the order of $(E_{-}^{III})^{2l+2}$ and $E_{-}^{III}$ respectively. So, the approximation has almost no effect in amplification factor, while the accuracy of quasinormal modes are always higher than $10^{-2}$, even $10^{-6}$ when $a\gtrsim 0.3$, provided with the regular parameter $k\le 0.01M$.

\end{appendix}

\end{document}